\documentclass[twocolumn,showpacs,amsmath,amssymb,prc,aps,10pt]{revtex4}

\usepackage{graphicx}
\usepackage{float}
\usepackage{dcolumn}
\usepackage{bm}
\usepackage{multirow}
\usepackage{amssymb}
\usepackage{amsmath}
\usepackage[usenames,dvipsnames]{xcolor}
\usepackage{doi}

\newcommand{\bb}{\color{blue}}

\begin{document}
\newcommand{\MeV}{\ensuremath{\,{\rm MeV}}}
\newcommand{\GeV}{\ensuremath{\,{\rm GeV}}}
\newcommand{\TeV}{\ensuremath{\,{\rm TeV}}}
\newcommand{\req}[1]{Eq.\,\ref{#1}}
\newcommand{\rf}[1]{Fig.\ref{#1}}

\topmargin -1.0cm
\title{Fusion reactions initiated by laser-accelerated particle beams\\ in a laser-produced plasma}

\author{C. Labaune$^1$}
\author{C. Baccou$^1$}
\author{S. Depierreux$^2$}
\author{C. Goyon$^2$}
\author{G. Loisel$^1$}
\author{V. Yahia$^1$}
\author{J. Rafelski$^3$}
\affiliation{%
$\quad$\\ 
$^1$LULI,~Ecole~Polytechnique,~CNRS,~CEA,~UPMC,~F-91128~Palaiseau,~France
}%
\affiliation{
$^2$CEA, DAM, DIF, F-91297 Arpajon,  France 
}%
\affiliation{
$^3$Department of Physics, The University of Arizona, Tucson, Arizona 85721, USA  \\ \phantom{[0.9cm]}
}%

\date{Received 24 Jan 2013, Accepted 27 Aug 2013}

\begin{abstract}
\vskip -6.4cm 
\begin{center}{\large\bf  \underline{\bb   Nature Communications 4:2506 \doi{10.1038/ncomms3506} (2013)}}\\ 
\end{center}
\vskip 5.4cm 
The advent of high-intensity pulsed laser technology enables the generation of extreme states of matter under conditions that are far from thermal equilibrium. This in turn could enable different approaches to generating energy from nuclear fusion. Relaxing the equilibrium requirement could widen the range of isotopes used in fusion fuels permitting cleaner and less hazardous reactions that do not produce high energy neutrons. Here we propose and implement a means to drive fusion reactions between protons and boron-11 nuclei, by colliding a laser-accelerated proton beam with a laser-generated boron plasma. We report proton-boron reaction rates that are orders of magnitude higher than those reported previously. Beyond fusion, our approach demonstrates a new means for exploring low-energy nuclear reactions such as those that occur in astrophysical plasmas and related environments.
\end{abstract}



\maketitle

\section{\label{sec:intro}Introduction}
Inertial confinement fusion research over the past 40 year has been focused on the laser-driven reaction of deuterium ($d$) and tritium ($t$) nuclei under near thermal equilibrium conditions~\cite{Nuckolls72,Lindl95}. The $dt$ fusion reaction was chosen because of its higher thermal reaction rate compared  to that of other light isotopes and a sustained burn can be achieved at relatively low temperatures ($\simeq 20$keV; Ref.~\cite{Atzeni04}). However, this reaction produces an intense flux of high energy neutrons ($n)$, which represents a significant radiation hazard and generates nuclear waste. Recent advances in laser technology~\cite{StrickMourou85}, laser-plasma interaction physics~\cite{Pesme93}, and laser-accelerated particle beams~\cite{Fews94,Umstadter00,Snavely00,Fuchs06} could enable the development of fuels based on aneutronic nuclear reactions that produce substantially less  radiation~\cite{FuelReview00,Nevins98}. In the case of the fusion reaction of protons ($p$) and boron-11 ($^{11\!}$B)  nuclei, fusion energy is released predominantly in the form of charged alpha ($\alpha$) particles~\cite{Becker87} rather than neutrons.  Moreover, boron is both, more plentiful than tritium, and easy to handle. At high temperature, the equilibrium thermal fusion rate of   $p^{11\!}$B is comparable to the $dt$-fusion rate~\cite{Nevins00}. However, the use of $p^{11\!}$B  with the spherical laser compression scheme would require excessively high laser energies to reach the high temperature and density necessary to achieve in thermal equilibrium fusion burn. Moreover, energy losses to Bremsstrahlung radiation under such conditions would prevent this reaction from being self-sustaining~\cite{Moreau77,pBEliezer96}. 

Such problems could be overcome by driving the  $p^{11\!}$B  reaction under conditions far from equilibrium, over shorter timescales than those involved in conventional inertial confinement fusion schemes, using short-pulsed high-intensity lasers. Laser-driven nuclear reactions are a new domain of physics~\cite{Ledingham10}, which aside from energy generation, are of interest to furthering the understanding of stellar nuclear processes~\cite{Adelberger11,Nomoto06} and of Big Bang nucleosynthesis~\cite{Coc12}. The first demonstration of a laser-driven  $p^{11\!}$B  reaction~\cite{Belyaev05} used a picosecond laser pulse at an intensity of  $2\times 10^{18}\mathrm{W/cm}^2$  focused onto a composite target $^{11}$\!B+(CH$_2)_n$1  resulting in $\simeq 10^3$   reactions, or more~\cite{Kimura09}, in $4\pi$ steradians.  The observed reaction yield was interpreted as the result of in-situ high-energy ions accelerated by high-intensity laser pulses~\cite{Gitormer86}. Other theoretical and numerical schemes have been considered in the past 15 years with the aim of realizing a $p$B fusion reactor: a colliding beam fusion device~\cite{Rostocker97}, fusion in degenerate plasma~\cite{Son04}, plasma block ignition driven by nonlinear ponderomotive forces~\cite{Hora09}, and proton pulse from Coulomb explosion hitting a solid B target~\cite{CoulombExpl11}.   In all these schemes one seeks to improve the expected ratio of energy gain to loss by various means, but none of them has come close to achieving this goal.

Here we demonstrate an approach that realizes a substantial increase in the rate of a laser-driven $p^{11\!}$B reaction. We achieve this by using  two laser beams. The first is a high energy, long pulse duration (nanosecond regime) laser that is focused on a solid target to form an almost completely ionized boron-11 plasma ($T_e\ge 0.5$\,keV). The second beam is a high intensity ($6\times 10^{18}\mathrm{W/cm}^2$), short pulse duration (picosecond regime) laser capable  of accelerating a high energy proton beam (see appendix \ref{Methods}). The picosecond timescale of a laser generated high intensity proton beam limits the ensuing radiation losses. Directing this beam into the plasma results in collisions with boron ions~\cite{Patent12}  at energies near to the nuclear resonance energies  of $E_p=162$ and $675$\,keV (for  $p +  ^{11}$\!B   resonances, see table 12.11 in Ref.~\cite{NPB90}).  Unlike the other efforts we do not wish to address an immediate potential for realization of an actual practical device, but to demonstrate scientific progress towards aneutronic fusion with short pulse lasers, and to present opportunities for future continuation of this research objective.

\section{Experimental set-up}
\subsection{Two laser beams at LULI2000}
Experiments have been carried out on the Pico2000 laser facility at the LULI laboratory. This installation synchronizes two laser beams as described above, for use in the same vacuum chamber. The long high-energy pulse delivers 400\,J in [1.5-4]ns, square pulse, at 0.53\,$\mu$m of laser light. It was focused with an f/8 lens through a random phase plate producing a focal spot diameter around 100\,$\mu$m (full width at half maximum) and an average intensity of $5\times 10^{18}\mathrm{W/cm}^2$. It was used to produce a plasma from a natural boron target (20\% of $^{10}$B and 80\% of $^{11}$\!B) placed at an incidence of 45$^0$ from the laser pulse propagation axis. The boron plasma expanded in vacuum producing an electron density profile from zero to solid $(\simeq 6\times 10^{23}\mathrm{cm}^{-3})$. The short laser pulse delivered 20\,J in 1\,ps with high contrast at 0.53\,$\mu$m wavelength. It was tightly focused on target reaching intensities $\simeq 6\times 10^{18}\mathrm{W/cm}^2$  to produce a proton beam by the TNSA (Target Normal Sheath Acceleration) mechanism~\cite{Passoni04}. Thin foils of aluminum, plastic and plastic covered by a thin layer of gold were irradiated at normal incidence. Details of the set-up are shown in \rf{Figure1}. The two beams were set at a relative angle of 112.5$^0$ from each other. The distance between the thin foil and the boron target was 1.5\,mm. The time delay between the two beams was adjusted between 0.25 and 1.2\,ns, so that the proton beam interacted with a plasma state in various conditions of ionization and temperature. Shots were done with either the short pulse only so the proton beam interacted with solid boron, or with the two laser beams so the proton beam interacted with boron plasma. 
\begin{figure}[!tb]
\includegraphics[width=\columnwidth]{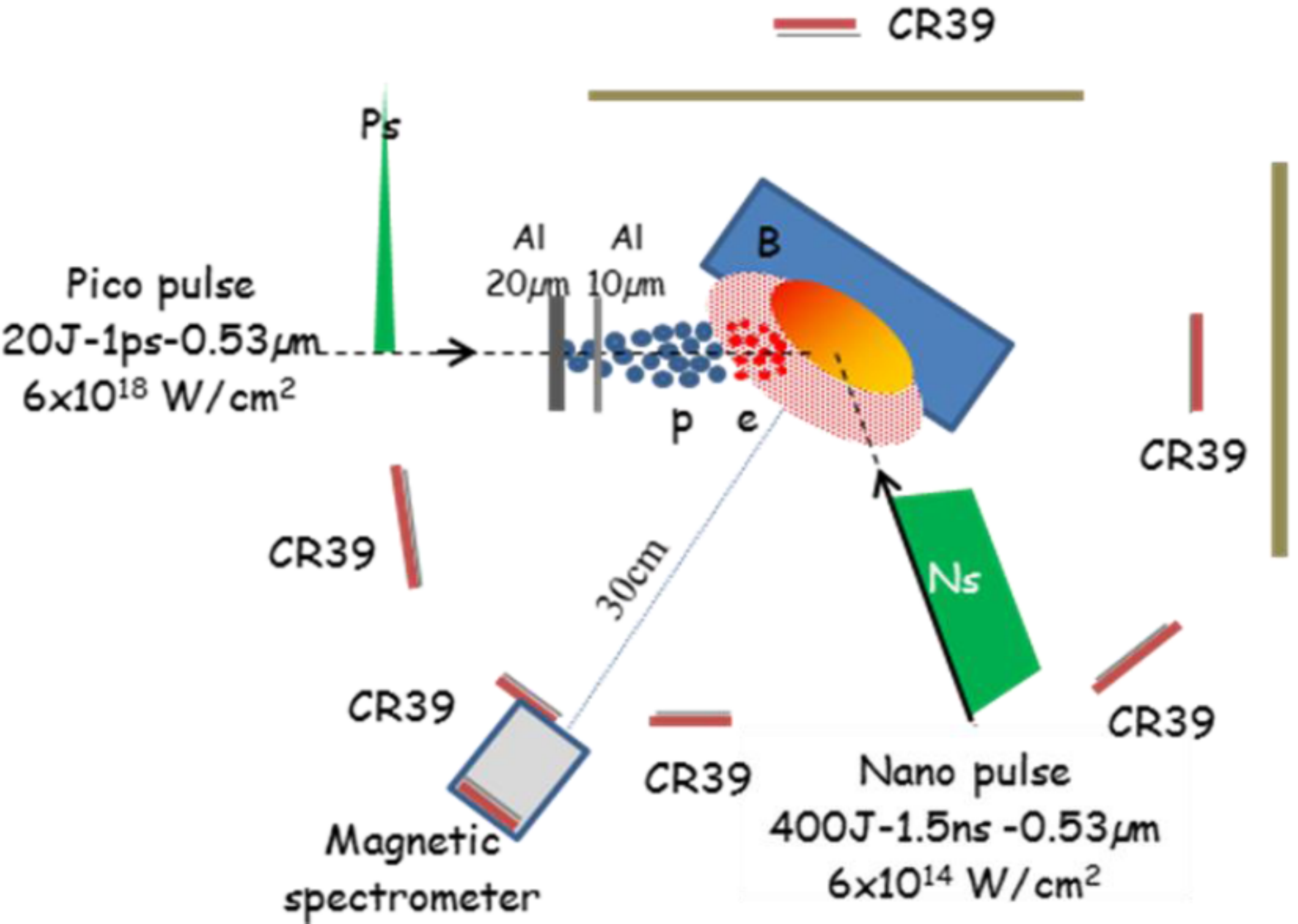}
\caption{\label{Figure1}: {\bf Experimental set-up.} Scheme of the experimental set-up showing the laser beam configuration, the target arrangement and the diagnostics (CR39 track detectors and a magnetic spectrometer). The picosecond pulse arrives from the left and generates a proton beam in the first 20\,$\mu$m Al foil, which impacts the boron plasma produced by the nanosecond pulse arriving from the bottom. The second 10\,$\mu$m Al foil protects the first one from irradiation by the nanosecond beam.}
\end{figure}

\subsection{Diagnostics}
Track detectors CR-39,~\cite{CR3907}, covered by aluminum foils of various thickness between 6 and 80\,$\mu$m were used to collect impacts by both protons and $\alpha$-particles. Six detectors were used for each shot with angles 0, 15, 35, 70, 100 and 170$^0$ from the picosecond beam axis (0$^0$ is the forward direction). A magnetic spectrometer, with a magnetic field of 0.5T, was placed along the normal of the boron target to analyze the $\alpha$-particle spectrum. An aluminum filter with 12\,$\mu$m thickness was placed in front of the slit of the spectrometer to block low-energy ions, Boron below 11\,MeV, Carbon below 14\,MeV, Oxygen below 19\,MeV and Aluminum below 23\,MeV. The tracks observed inside the spectrometer are therefore mainly ascribable to protons and $\alpha$-particles which impact at the same position when having the same entrance point and same energy inside the spectrometer. Taking into account the loss of energy in the aluminum filter at the entrance of the spectrometer, $\alpha$-particles with energy between 3.3 and 7.5\,MeV and protons between 0.9 and 5\,MeV could be measured. This is also the method by which we characterized the proton beam spectra in preparation for fusion shots.

\subsection{Plasma characterization}
In the main part of the experiment, the objective was to study the number of $p^{11}$\!B reactions between the proton beam accelerated by the picosecond laser and the boron for different prepared target conditions. The expansion of the boron plasma produced by the nanosecond pulse was characterized by time-integrated X-ray pinhole images in the range $\simeq$3-5\,keV. Typically the overall extension of the boron plasma was around 200\,$\mu$m after 1\,ns. This diagnostic was also very useful to control the alignment and superposition of the two beams as shown in \rf{Figure2} where three plasmas are observed along the direction of propagation of the picosecond beam: the first one comes from the 20\,$\mu$m Al foil which is used to produce the proton beam, the second one comes from a second 10\,$\mu$m Al foil that was inserted to protect the rear part of the first Al foil from the nanosecond scattered photons by the boron target and the third one is the boron plasma. Without the 10\,$\mu$m Al shield, the proton beam could not be produced in the two-beam irradiation shots because light scattered from the nanosecond pulse modified the rear surface of the Al foil, and as it is believed, cleaning up all the hydrogen rich impurities~\cite{pBeam08}. Finally, an estimate of the electronic temperature of  the boron plasma was obtained from the shift of the time-resolved stimulated Brillouin backscattering (SBS) spectra of the nanosecond pulse~\cite{Kruer03}. They were recorded with a high-dispersion spectrometer and a streak camera. A typical example of such time-resolved SBS spectrum is shown in \rf{Figure3} in the case of a 4\,ns pulse irradiating the boron target. The spectral shift of the SBS light was analyzed using the ion-acoustic velocity formula, providing an electron temperature of $T_e \simeq(0.7\pm0.15)$\,keV.
\begin{figure}[!tb]
\includegraphics[width=0.8\columnwidth]{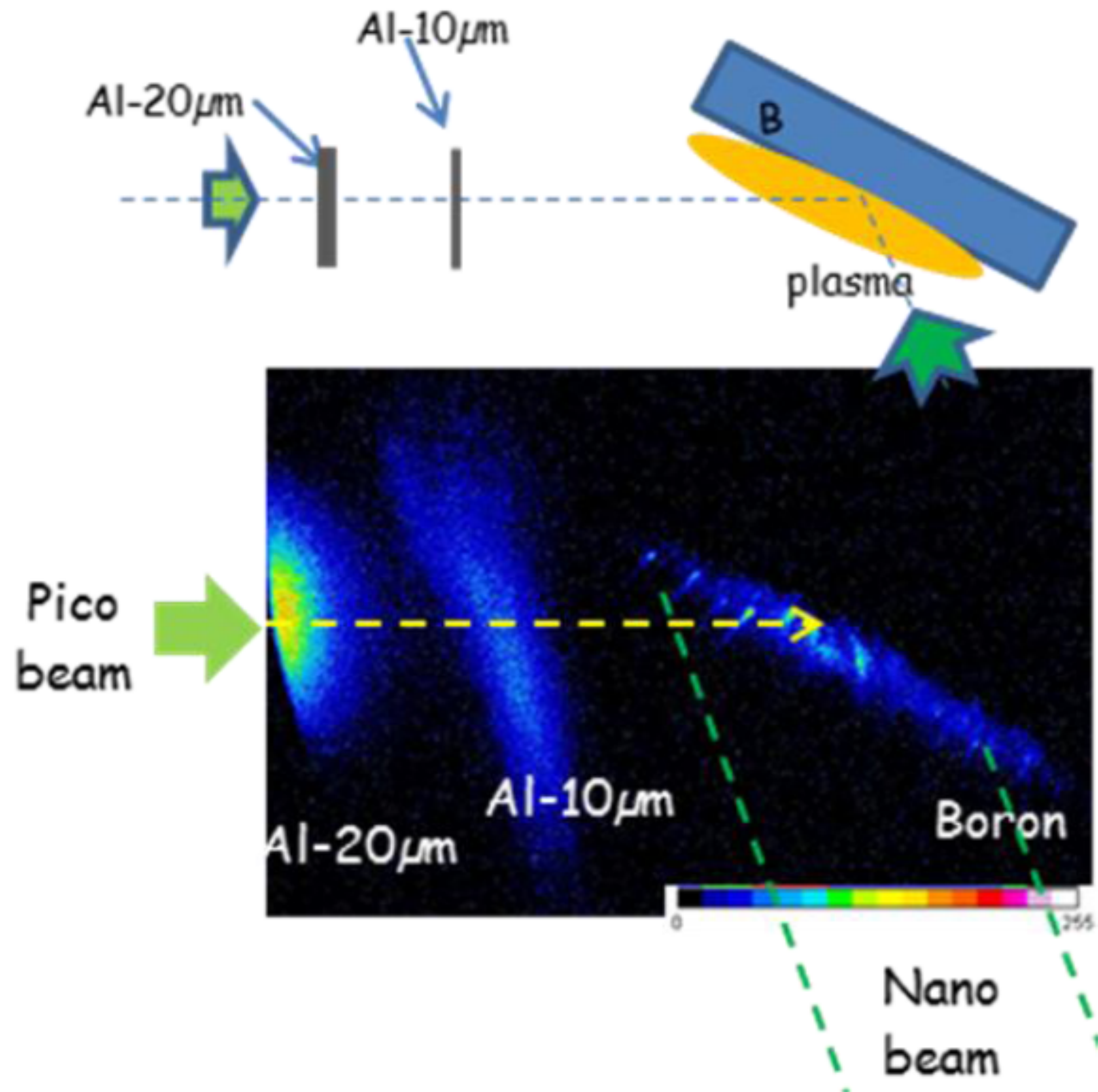}
\caption{\label{Figure2}: {\bf Observation of the multiple plasmas.} Time-integrated X-ray pinhole image of the three plasmas along the direction of propagation of the picosecond beam. From left to right, we observe the heated parts of the first Al foil that produces the proton beam, the second Al foil that protects the first one and the boron plasma.}
\end{figure}

\begin{figure}[!tb]
\includegraphics[width=0.75\columnwidth]{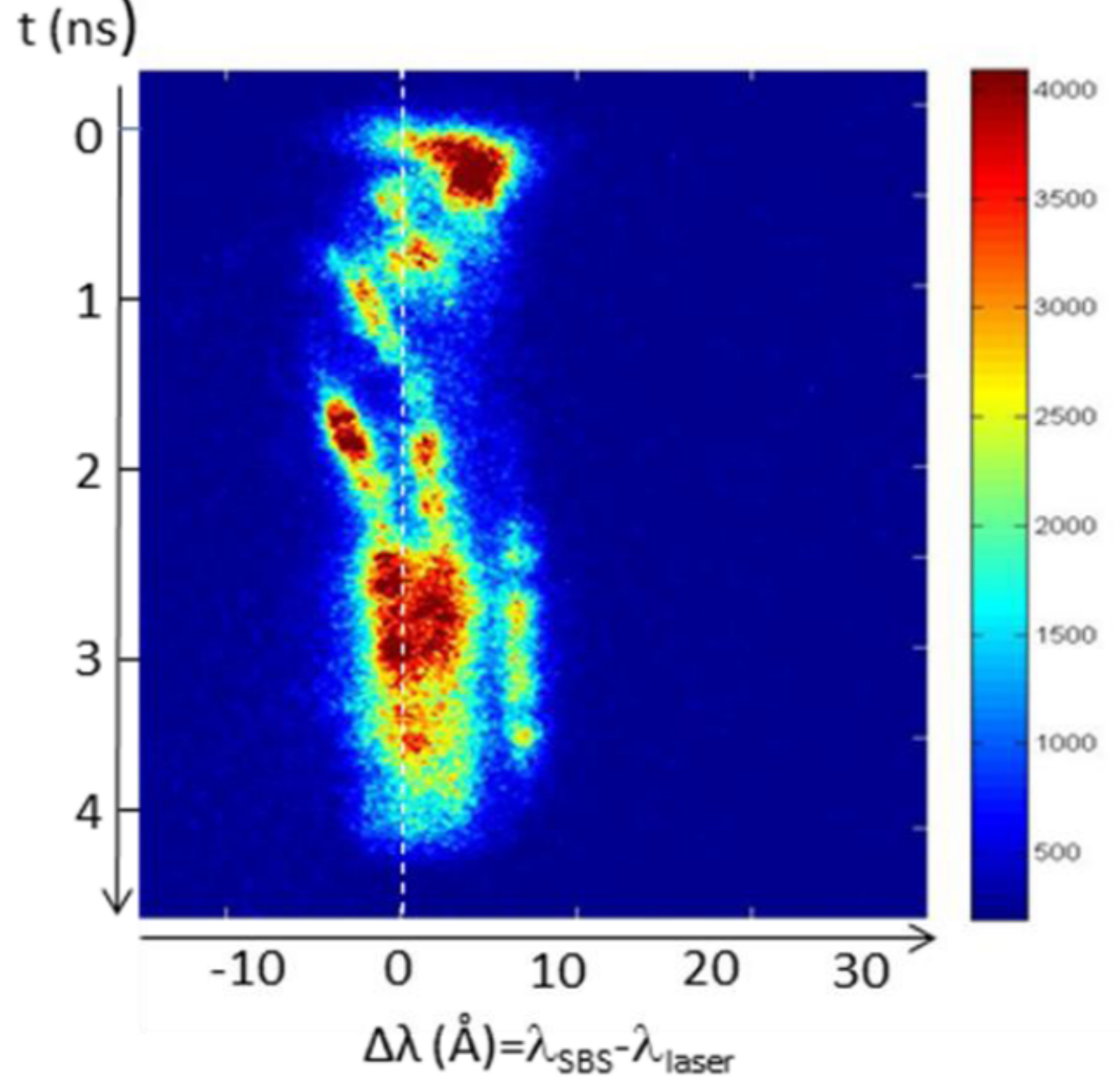}
\caption{\label{Figure3}: {\bf Stimulated Brillouin scattering spectrum.}  Time-resolved spectrum of stimulated Brillouin backscattering (SBS) of the nanosecond pulse from the boron plasma. The light is collected in the focusing optics of the nanosecond beam in the backward direction. The laser pulse is at 0.53\,$\mu$m with an intensity of $5\times 10^{14}\!\mathrm{\,W/cm}^2$.}
\end{figure}

\section{Experimental results}
The total number of tracks observed in the magnetic spectrometer, per unit of surface on the CR39, as a function of the $\alpha$-particle energy is shown in \rf{Figure4} for various shot conditions. No particle can be observed in the hatched part owing to the aluminum filter in front of the spectrometer. Shots with no boron target behind the Al foil (yellow triangles) display almost no track which means that very few protons are accelerated at an angle of 100° from the pico beam axis as expected from the TNSA process. In the case of shots with the picosecond beam alone, in which the proton beam interacts with a solid boron (blue diamonds), the number of tracks is close to the noise level, indicating very weak activity. Shots with the two beams, in which the proton beam interacted with boron plasma, demonstrate a large increase of the number of tracks by a factor of more than a hundred, in the highest case, compared to the previous ones. Three time-delays between the two beams have been tried: 0.25\,ns (open circles), 1\,ns (green triangles) and  1.2\,ns (blue squares) showing that the highest number of tracks was obtained for the longest time-delay which corresponds to the highest temperature and ionization state of the boron plasma.
\begin{figure}[!tb]
\includegraphics[width=0.9\columnwidth]{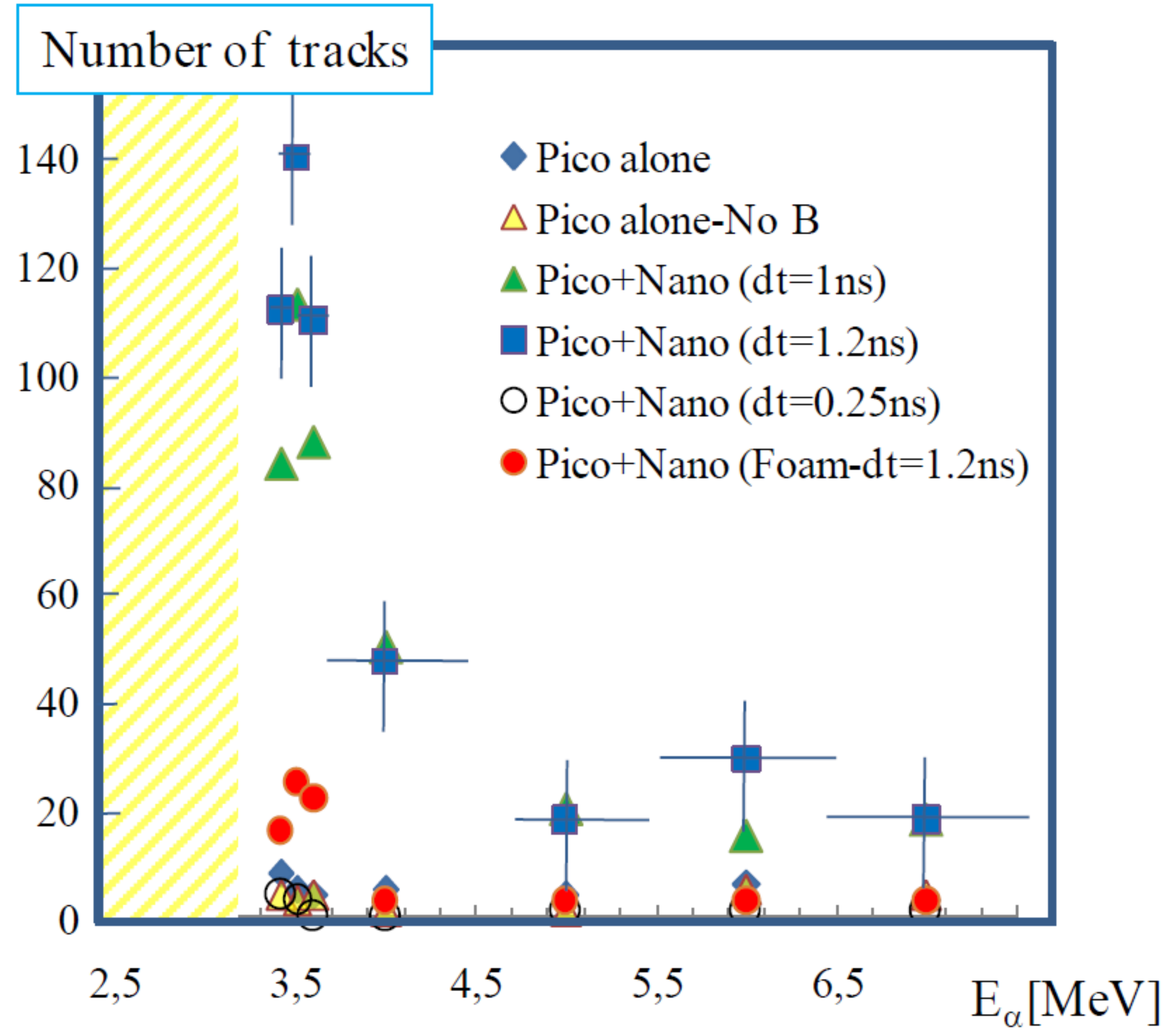}
\caption{\label{Figure4}: {\bf $\alpha$-particle spectra.} The total number of tracks observed in the magnetic spectrometer (with an entrance slit of 1\,mm$^2$) per unit of surface of CR39 as a function of the $\alpha$-particle energy for six shot configurations: yellow triangles = shot with no boron; blue diamonds = interaction of the proton beam with solid boron; blue square, green triangles = interaction of the proton beam with plasma boron and time-delay between the two beams of 1 and 1.2\,ns respectively; red circles = ibid, where the proton beam is produced in a foam rather than aluminum foil; open circles = short delay (0.25\,ns) between the nano and the pico pulses. The error bars in energy are given by the width of the CR39 on which the number of impacts has been counted; the error bars in the number of impacts are given by the shot to shot fluctuations ($\simeq\pm10\%$). The low energy domain has no counts as the entrance slit is protected by a 12\,$\mu$m Al foil.}
\end{figure}

These results were complemented by the analysis of the CR39 detectors which were positioned outside the spectrometer, close to the entrance slit. Tracks were observed only behind aluminum filters of thickness smaller than 24\,$\mu$m. If scattering of the proton beam by the boron plasma had sent protons into the spectrometer, tracks would have been recorded for all the aluminum thicknesses as the proton beam includes a continuous spectrum of energy up to $\simeq 10$\,MeV (see appendix \ref{Methods}) which can cross a thickness of aluminum larger than 80\,$\mu$m. This is not the case for the produced $\alpha$-particles. Given our proton spectrum, the kinetic energy range of $\alpha$-particles produced in $p^{11}$\!B reactions~\cite{Becker87} is 0.5-8\,MeV.  Considering the exponential decrease of proton yield with energy there is little if any production of $\alpha$-particles with energy larger than 7.1\,MeV required to cross 36\,$\mu$m or more of aluminum. $\alpha$-particles with typical fusion energy between 3.3 and 5.4 MeV can cross 12 and 24\,$\mu$m of aluminum as observed in the $p^{11}$\!B shots. To conclude, the absence of high energy proton signature in control detectors placed near to the spectrometer is our evidence that scattered protons are not producing the track signature inside the spectrometer.

A rough estimate of the fusion rate can be obtained from the number of tracks in the spectrometer and the solid angle of observation ($\delta\Omega=1.1\times 10^{-5}$ sr). The highest event rate measured in this scheme was $9\times 10^6$/sr which is much higher than previous observations~\cite{Belyaev05}. However, as pointed out in Ref.~\cite{Kimura09}, the choice of the detection energy region of the reaction products can underestimate the total yield as $\alpha$-particles with energy lower than 3.3\,MeV are not taken into account. In our experimental conditions, the low energy $\alpha$-particles are not expected to escape the plasma and furthermore those having relatively small energy when escaping may not be observed leading to an underestimate of the absolute fusion yield~\cite{Kimura09}. External $\alpha$-particle detectors can only observe fusion products emitted within the plasma in a backward hemisphere at an energy allowing escape from the plasma. This means that $\alpha$-particles propagating in the forward direction, into the thick solid target, cannot be observed directly. These are accounted for by consideration of the solid angle of observation of the spectrometer. Test shots were performed with either the nano or the pico pulse alone on the boron target to measure the possible reactions in the hot plasma and the number of tracks in both cases was below 10. This demonstrates that the observed high number of particles in the two-beam experiments is definitely the consequence of the interaction of the proton beam with the boron plasma.

\section{Discussion}
The features of the energy spectrum of $\alpha$-particles presented in \rf{Figure4} agree well with the $p^{11}$\!B fusion spectrum. The rise at $E<5$\,MeV is a well known feature of the spectra arising from the formation of the broad $^8\mathrm{Be}^*$ resonance in first step with the subsequent $^8\mathrm{Be}^*\to \alpha+\alpha$ decay products seen both experimentally~\cite{Stave11}  and understood theoretically~\cite{Dimitriev09}. The indication of a drop in the spectrum at the edge of our sensitivity near 3.5\,MeV could be the result of reaching the spectrometer edge, but is also a  $p^{11}$\!B fusion feature observed in other experiments, and expected theoretically. The small bump near to 6\,MeV may correspond to the two-body reaction $p + ^{11}\!\mathrm{B}\to  ^8\mathrm{Be} + \alpha + (8.59+E^*)$\,MeV,  at the reaction resonance energy $E^*$, Ref.~\cite{NPB90}, where 2/3 of the available energy (6--6.5\,MeV ) is carried away by the $\alpha$-particle. This bump is experimentally observed in thin target experiments~\cite{Stave11} but, considering our measurement error bars, is not a compelling feature in our results. Overall, the $\alpha$-particle spectrum in \rf{Figure4}  can be explained by the main known characteristic features of the  $p^{11}$\!B fusion.

There are several possible plasma state mechanisms modifying the $p^{11}$\!B fusion yield. Recall that protons entering a solid atomic target use most of their energy to ionize atoms and do not penetrate beyond a thin layer on the front surface. In the case of a preformed plasma, energetic protons ($>0.5$\,MeV) can be subject to reduced stopping power~\cite{Inject00}, and so penetrate deeper inside the plasma. Moreover, since we are employing the TNSA mechanism to produce the proton pulse, we know that the proton beam is Coulomb-pulled by a relativistic electron pre-pulse cloud. This cloud contains around 10-30\% of the pico pulse energy and impacts the boron plasma about 100\,ps ahead for our geometry: the distance between the thin foil and the boron target was 1.5\,mm, which at the velocity of light, corresponds to 5\,ps travel time. In comparison, a proton of kinetic energy $E_p=1$\,MeV and velocity $(v_p/c)^2= 2E_p/(m_p c^2)$ will take 21.7\,times longer, that is $108$\,ps to travel this distance. The relativistic electron cloud may condition the boron plasma just in the proton-target area, pushing out the electrons and forming an ionic channel. The consequence is that the proton pulse energy loss caused by interactions with plasma electrons is reduced while number of interactions with boron atomic nuclei is correspondingly enhanced. Note that because of the proton-boron mass asymmetry, protons lose relatively little energy in each deep near-nuclear Coulomb collision. Therefore across the large width $\simeq 250$\,keV of the $E_p=675$\,keV (proton energy $E_p$ on rest boron target) resonance~\cite{NPB90}, many such interactions can result in a large probability of fusion yield per proton. Furthermore the fusion cross section may be modified by plasma effects e.g. a modified electron screening~\cite{Angulo93,Barker02,Kimura04}  of the boron nuclei which may not be completely ionized. However this effect so far has been observed to be significant only at lowest reaction energies but is not well-understood in hot plasmas.

The rate of proton initiated fusion in a $^{11}$\!B target is $\lambda_f = \sigma_f\,\rho_B\, v_r$, where $v_r$ is the relative $p$ -- $^{11}$\!B  velocity,  $v_r=c/26$ -- $c/33$ in $E_p =0.68$ -- $0.43$\,MeV proton energy domain of interest, $\rho_B$ is $^{11}$\!B target density, which we take as a fraction of the solid natural target  $\rho_{0B} = 1.0\times 10^{23}$\,atoms/cm$^3$, and $\sigma_f$ is the resonant fusion cross section which averages in the interval of energy of the proton  $E_p=0.68$ -- $0.43$\,MeV to 1 barn~\cite{Nevins98}. This gives an average fusion reaction rate of $\lambda_f = 1/(100\mathrm{ns})$. In our situation the number of fusions achievable per proton is limited by the active depth of the reduced density plasma target which protons will traverse in $\simeq (1/30)$\,ns. Allowing for available higher energy protons in laser generated particle beam, in our present experimental conditions, we expect about 1 in 300 -- 3000 protons will be able to induce a $p^{11}$\!B  fusion reaction. Assuming that $10^{-3 }$ of the protons produce a $p^{11}$\!B   fusion reaction, the total number of reactions can be estimated very crudely by $N=n_1\,n_2\,\sigma \,v$, which gives $N\simeq 8\times 10^7$ in $4\pi$, (with $n_1=5\times 10^9$, $n_2=4\times 10^{14}$, assuming a reacting volume of $4\times 10^{-8}$\,cm$^3$, cylinder of 20\,$\mu$m radius and 30\,$\mu$m length, and an average density of $10^{22}$cm$^{-3}$). This corresponds to 88 $\alpha$-particles in the solid angle of observation ($\delta \Omega=1.1\times 10^{-5}$\,sr), in qualitative agreement with the observed numbers. 

Concerning the observability of $\alpha$-particles produced in fusion, it is clear that it depends on their energy spectrum and angular diagram of emission, which then depends on the relative cross sections of the different possible reactions~\cite{Becker87,Kimura09}, which are unknown under our conditions. If for some reason there are some differences in fusion reactions in solid compared to plasma medium, either in the effective cross section or in their capability to escape from the target, this could contribute to the modification in observed $\alpha$-count rates. Future experimental work will be dedicated to test the relative importance of these different mechanisms. 

Although our results are specific to the  $p^{11\!}$B   case, a similar approach could be used to study the reaction of other light isotopes. This provides a new approach to exploring aneutronic nuclear fusion reactions in dense plasma environment. Furthermore, our method could enable progress in the development of so-called fast-ignition fusions scheme~\cite{Roth01} by providing a short lived hot spot generated by both the particle beam and the $\alpha$-particles produced in the reactions, initiating and promoting a propagating burn wave.  The $p^{11\!}$B   case is unique in that secondary $\alpha$B reactions can regenerate the high energy proton, sustaining a fusion chain. Our experimental approach also suggests opportunities to explore nuclear reactions of astrophysical interest~\cite{40} in an environment more similar to the early universe or stellar interiors.

\subsection*{Acknowledgments}
We acknowledge the support of the LULI teams, and discussions with V. Tikhonchuk,  W. Rozmus. JR wishes to thank Ecole Polytechnique for the support of his three months sabbatical research visit in 2012, and Christine Labaune for hospitality at the LULI laboratory where this work was carried out.

\subsection*{Contributions}
C.L., G.L., and V.Y. were involved in the experimental project planning and the target design and carried out the experiments. S.D. and C.G. participated in the experiment and were involved in the optimization of the multiple targets configuration. C.B. implemented, developed, and analyzed the CR39. C.L. and J.R. conceived the idea of the scheme, were leaders of the analysis team and wrote jointly all research reports including this one. 
  
\appendix
\section{\label{Methods}Methods}
{
{\bf Set up:} Experiments were conducted on the LULI2000 laser installation at Ecole Polytechnique. Two beams are produced by synchronized oscillators and amplified in similar neodymium glass chains. The first one delivers nanosecond pulses of 1\,kJ and the second one delivers picosecond pulses of 100\,J which are stretched before amplification and then compressed before focusing using the CPA (chirped pulse amplification) method. Both beams are initially at wavelength 1.06\,$\mu$m and then converted to the second harmonic just before focusing, which produces a high contrast for the pico pulse which is important for proton beam acceleration. CR39 detectors were etched after irradiation during 6--12 hours in a solution of NaOH in H$_2$O at 70\,C. The number and diameters of the tracks were analyzed using a Nikon microscope with a magnification of 20. The calibration of the CR39 for $\alpha$-particles was done using a $^{233}$U source which delivers $\alpha$-particles with energy of 5\,MeV. By increasing the thickness of the air layer between the source and the CR-39, lower energy $\alpha$-particles could also be observed.}

{
{\bf Laser-accelerated proton beam:} Protons with the required high kinetic energy are now routinely produced by short laser pulses having intensity on target higher than $2\times 10^{18}\mathrm{W/cm}^2$. Therefore, high-intensity lasers are a new tool in the study of nuclear fusion reactions in a high density regime. Their unique features are the formation of a high intensity proton pulse of time duration similar to the laser pulse, and a tunable spectral distribution. 

The first part of the experiment was dedicated to the proton beam optimization. In their interaction with the boron target, protons with energy around 170\,keV and 700\,keV would be most capable to take direct advantage of the resonances in the cross section of the  $p^{11}$\!B reactions. Nevertheless, we believed that in our target condition a broad energy spectrum and high intensity proton pulse had a greater advantage to increase the fusion yield.  Our boron target was thick and the electron density profile of the boron plasma created by the nano-pulse displayed all densities up to the solid assuring that the incoming particle pulse would be stopped. So, we chose to optimize the number of above 1\,MeV protons in the particle pulse. We tried out three types of targets: CH with 2\,$\mu$m thickness covered by 125\,nm of gold, low-density (3mg/cc) cellulose-triacetate-(C$_{12}$H$_{16}$O$_{8}$) TAC foams~\cite{TACfoam} with 300\,$\mu$m length and Al foils with thickness 10 and 20\,$\mu$m. The largest number of high-energy protons on the laser axis in the forward direction was obtained with the 20\,$\mu$m Al foils,the protons are known to originate from hydrogenated deposit on the back of the Al foil.}

\begin{figure}[!tb]
\includegraphics[width=0.9\columnwidth]{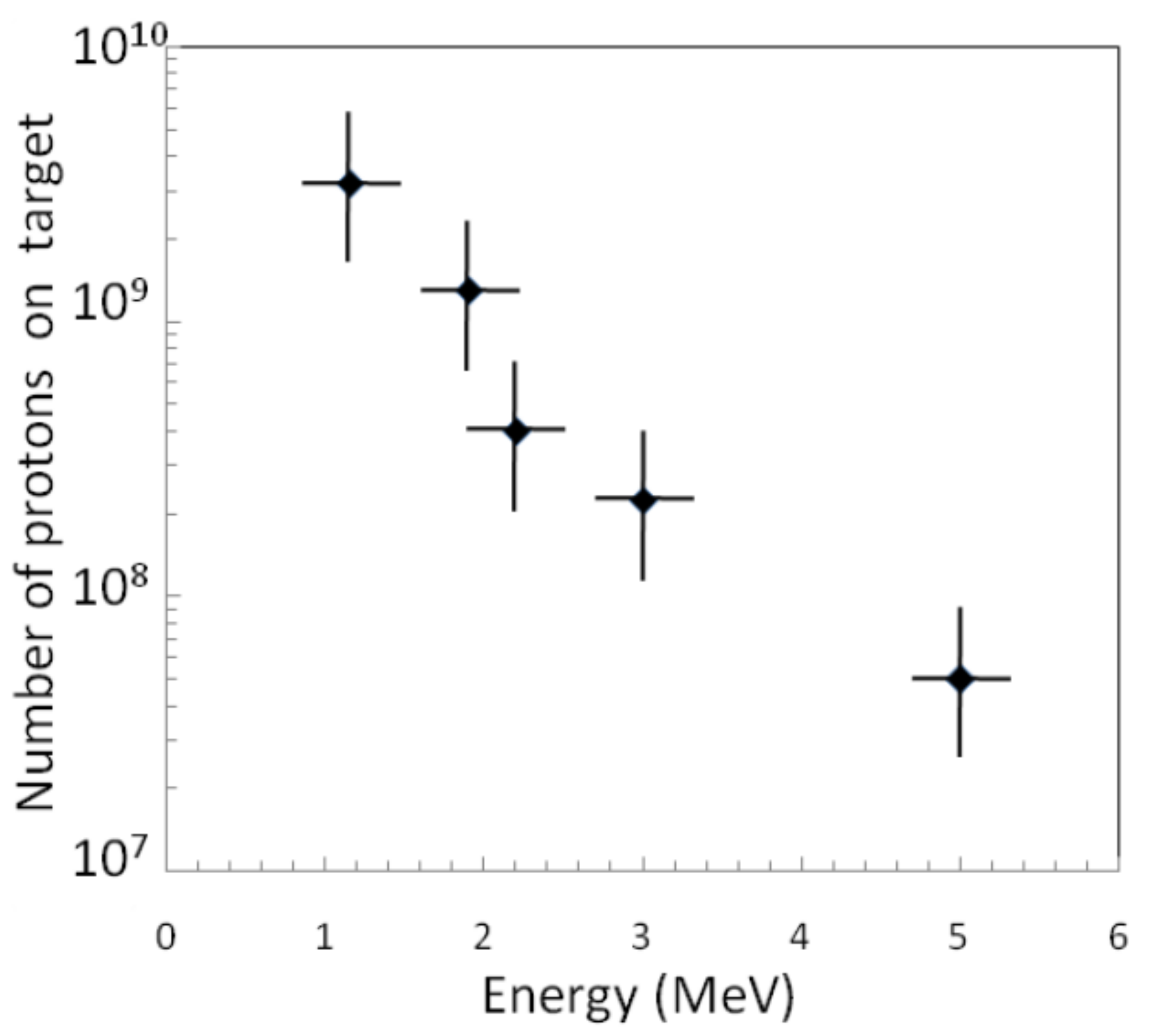}
\caption{\label{Figure5}: {\bf Proton spectrum.} Energy spectrum of the proton beam produced by the interaction of an aluminum foil of 20\,$\mu$m thickness with the LULI2000 picosecond pulse at 0.53\,$\mu$m, with a pulse duration of 1\,ps and an intensity of  $6\times 10^{18}\!\mathrm{\,W/cm}^2$. The error bars in energy are given by the differences in thickness of the Al filters covering the CR39 on which the number of impacts has been counted; the error bars in the number of impacts are given by the shot to shot fluctuations  ($\simeq\pm10\%$).}
\end{figure}

{
An estimate of the energy distribution of the proton beam in forward direction was obtained by analyzing the number of impacts on the CR-39 covered by 24, 36, 44, 56, 60, 72 and 80\,$\mu$m of aluminum. In addition, the absolute number of protons with energy higher than 5\,MeV was deduced from the boron activation which produces $^{11}$C through the nuclear reaction~\cite{Ledingham04}:   $p + ^{11}\!\mathrm{B} \to ^{12}\!\mathrm{C}\to  n + ^{11}\!\mathrm{C} -2.9$\,MeV. The $^{11}$C has a half-time decay of 20.334 minutes and was measured from the residues of the target just after the shot until one hour later. From these measurements, we deduced that per “fusion” shot more than $5\times 10^7$ protons with energy larger than 5 MeV were produced by the pico laser pulse and arrived on the target. We further determined that the angular emission of protons in the pulse was strongly peaked along the laser axis with a typical half-angle of $\simeq 5^0$. An example of the energy spectrum of the proton beam produced by the interaction of an aluminum foil of 20\,$\mu$m thickness with the picosecond pulse at an intensity of $6\times  10^{18}\,\mathrm{W/cm}^2$ is shown in \rf{Figure5}. A continuous spectrum of energy was observed up to $\simeq 10$\,MeV in agreement with other observations for similar conditions.}

{
{\bf Two-beam experiments:} The main goal of the experiments we report here was to demonstrate the effect of the preparation of the target boron plasma state on the observed reaction rate. The largest number of CR-39 associated tracks was observed in the case of the best geometric superimposition of the proton beam and the boron plasma, when the proton beam was produced from a 20\,$\mu$m thick Al foil and arrived close to the end of the nanosecond pulse, so the boron plasma was at maximum temperature and ionization. When using a 3\,mg/cc 300\,$\mu$m long foam to produce the protons, the number of tracks was reduced by $\simeq 7$ compared to the case where the proton beam was from a 20\,$\mu$m thick Al foil, with all the other parameters being the same (red circles in \rf{Figure4}). This reduction may be directly attributed to the reduction of the total number of MeV energy protons achieved in the case of the foam target. Complementary shots were dedicated to establish the mechanisms leading to our results, and of most interest in the present discussion is the case in which we fired the picosecond pulse on a target composed of boron covered by a 0.9\,$\mu$m CH foil to produce high energy ions in a plasma mixture of $p$ and B. The number of tracks was close to our noise level. Those conditions are close to the ones used by Belyaev~\cite{Belyaev05}  further demonstrating that the production of a particle pulse comprising high-energy protons (above 1 MeV) in a separate target could be the origin of  the significant increase  of the fusion yield.}


\end{document}